\documentclass[aps,prl,10pt,twocolumn,superscriptaddress]{revtex4-2}
\usepackage{amssymb}
\usepackage{eurosym}
\usepackage{dcolumn}
\usepackage{bm}
\usepackage{graphicx}
\usepackage[dvipsnames]{xcolor}
\usepackage{amsmath}
\usepackage{amsfonts}
\usepackage{lmodern}
\usepackage[colorlinks,linkcolor=black,citecolor=black,filecolor=black,urlcolor=Blue,bookmarksopen=false]{hyperref}
\usepackage{lipsum}
\usepackage{array}
\usepackage{afterpage}
\usepackage{orcidlink}

\begin{document}
\title{Competing Ordering Modes in the Distorted Quantum Kagome Material Clinoatacamite Cu$_2$Cl(OH)$_3$}
\author{L. St\"odter}
\email[Contact author:~]{l.stoedter@fz-juelich.de}
\affiliation{Institut f\"ur Physik der Kondensierten Materie, Technische Universit\"at Braunschweig, 38106 Braunschweig, Germany}
\affiliation{J\"ulich Center for Neutron Science (JCNS) at Heinz Maier-Leibnitz Zentrum (MLZ), Forschungszentrum J\"ulich GmbH, 85748 Garching, Germany}
\author{C. Kastner}
\affiliation{Institut f\"ur Physik der Kondensierten Materie, Technische Universit\"at Braunschweig, 38106 Braunschweig, Germany}
\author{H. O. Jeschke}
\affiliation{Research Institute for Interdisciplinary Science, Okayama University, Okayama 700-8530, Japan}
\author{M. Reehuis}
\affiliation{Helmholtz-Zentrum Berlin f\"ur Materialien und Energie GmbH, 14109 Berlin, Germany}
\author{E. Chan}
\affiliation{Institut Laue-Langevin, 38042 Grenoble Cedex 9, France}
\author{M.-H. Lemée-Cailleau}
\affiliation{Institut Laue-Langevin, 38042 Grenoble Cedex 9, France}
\author{K. Beauvois}
\affiliation{Institut Laue-Langevin, 38042 Grenoble Cedex 9, France}
\author{B. Ouladdiaf}
\affiliation{Institut Laue-Langevin, 38042 Grenoble Cedex 9, France}
\author{F. Yokaichiya}
\altaffiliation[Present address:~]{Department of Physics, Federal University of Parana, Curitiba, Parana, Brazil}
\affiliation{Helmholtz-Zentrum Berlin f\"ur Materialien und Energie GmbH, 14109 Berlin, Germany}
\author{F. Bert}
\affiliation{Universit\'e Paris-Saclay, CNRS, Laboratoire de Physique des Solides, 91405 Orsay, France}
\author{T. J. Hicken}
\affiliation{PSI Center for Neutron and Muon Sciences, 5232 Villigen PSI, Switzerland}
\author{J. A. Krieger}
\affiliation{PSI Center for Neutron and Muon Sciences, 5232 Villigen PSI, Switzerland}
\author{H. Luetkens}
\affiliation{PSI Center for Neutron and Muon Sciences, 5232 Villigen PSI, Switzerland}
\author{J. L. Allen}
\affiliation{Institute for Superconducting and Electronic Materials and School of Physics, University of Wollongong, NSW 2522, Australia}
\author{R. Feyerherm}
\affiliation{Helmholtz-Zentrum Berlin f\"ur Materialien und Energie GmbH, 14109 Berlin, Germany}
\author{M. Tovar}
\affiliation{Helmholtz-Zentrum Berlin f\"ur Materialien und Energie GmbH, 14109 Berlin, Germany}
\author{D. Menzel}
\affiliation{Institut f\"ur Physik der Kondensierten Materie, Technische Universit\"at Braunschweig, 38106 Braunschweig, Germany}
\author{A. U. B. Wolter}
\affiliation{Leibniz Institute for Solid State and Materials Research IFW Dresden, 01069 Dresden, Germany}
\author{B. B\"uchner}
\affiliation{Leibniz Institute for Solid State and Materials Research IFW Dresden, 01069 Dresden, Germany}
\affiliation{Institut f\"ur Festk\"orper- und Materialphysik and W\"urzburg-Dresden Cluster of Excellence ct.qmat, Technische Universit\"at Dresden, 01062 Dresden, Germany}
\author{K. C. Rule}
\affiliation{Australian Nuclear Science and Technology Organisation, Lucas Heights, NSW 2234, Australia}
\author{F. J. Litterst}
\affiliation{Institut f\"ur Physik der Kondensierten Materie, Technische Universit\"at Braunschweig, 38106 Braunschweig, Germany}
\author{U. K. R\"o{\ss}ler}
\affiliation{Leibniz Institute for Solid State and Materials Research IFW Dresden, 01069 Dresden, Germany}
\author{S. S\"ullow}
\affiliation{Institut f\"ur Physik der Kondensierten Materie, Technische Universit\"at Braunschweig, 38106 Braunschweig, Germany}
\date{\today}

\begin{abstract}
We have studied the magnetic properties of clinoatacamite Cu$_2$Cl(OH)$_3$, the parent compound of the quantum spin liquid candidate herbertsmithite and a longstanding puzzle among frustrated quantum magnets. As we reveal using density-functional theory, clinoatacamite belongs to the class of distorted kagome antiferromagnets with the kagome plane being embedded into a low-symmetry crystal structure. By means of thermodynamic measurements, muon spin rotation/relaxation as well as neutron diffraction on single crystals, we find in zero magnetic field complex behavior below $T_\mathrm{I} = 18.1$\,K which unfolds in three temperature regions I-III. We propose this complexity in multicritical clinoatacamite to arise from the competition of antiferromagnetic ordering modes from the underconstrained manifold of modes, which can lead to a metamagnetic texture in zero field.
\end{abstract}

\maketitle

For decades, experimental research on antiferromagnetic (AFM) quantum kagome materials has been primarily driven by the hunt for the quantum spin liquid (QSL) ground state~\cite{Balents2010,Broholm2020}. Yet, despite immense efforts, this highly entangled quantum spin state turned out to be hard to observe in real materials. Notably, the mineral herbertsmithite ZnCu$_3$(OH)$_6$Cl$_2$ stands out from this research on kagome magnetic insulators as a promising QSL candidate with the absence of magnetic order or spin freezing down to mK temperatures~\cite{Norman2016,Mendels2007,Khuntia2020}. This experimental observation motivated the research on an entire family of kagome materials as well as the synthesis of several new kagome compounds~\cite{Kermarrec2011,Boldrin2015,Puphal2018,Okuma2019,Henderson2019,Akazawa2020,Tustain2021,Chatterjee2023,Wang2023}. In effect, the years have shown that the sought-after QSL state is extremely unstable: Most $S=1/2$ kagome materials adopt static order instead of QSL behavior, and, to date, it has remained unclear what defines in detail the boundary between static order and highly correlated disorder.

More recently, another direction of kagome research has formed, which explicitly turns towards \emph{distorted} quantum kagome materials featuring a non-uniform motif of exchange couplings. It has been demonstrated that such a non-uniformity of the kagome lattice can lead to non-QSL but still exotic (quantum) behavior as, for instance, patterned valence bond solid states~\cite{Matan2010,Grbic2013}, coplanar order~\cite{Hering2022,Chatterjee2023}, trimer networks~\cite{Janson2016} and unusual applied-field properties such as magnetization plateaus~\cite{Matan2019}. Some of these properties have no counterpart in the uniform, ideal AFM kagome lattice described by the Hamiltonian 

\begin{equation}
\mathcal{H} = J\,\sum_{\left\langle i,j \right\rangle}\mathbf{S}_{i}\cdot\mathbf{S}_{j}.
\label{eq:hamiltonian}
\end{equation}

\noindent Here, $J$ denotes the exchange interaction between nearest-neighbor (NN) spins on the corner-sharing spin triangles and $\mathbf{S}_{i,j}$ are the spin operators at sites $i,j$. 

\begin{figure*}[t!]
\centering
\includegraphics[width=0.66\linewidth]{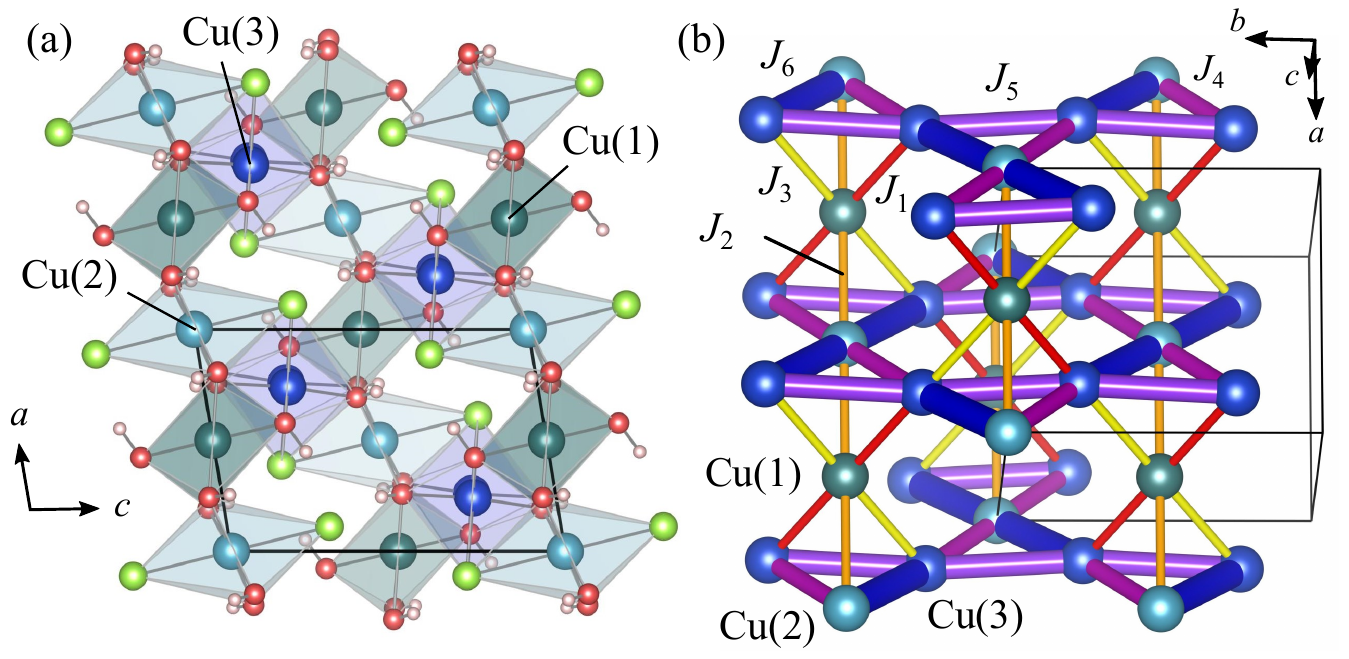}
\includegraphics[width=0.33\linewidth]{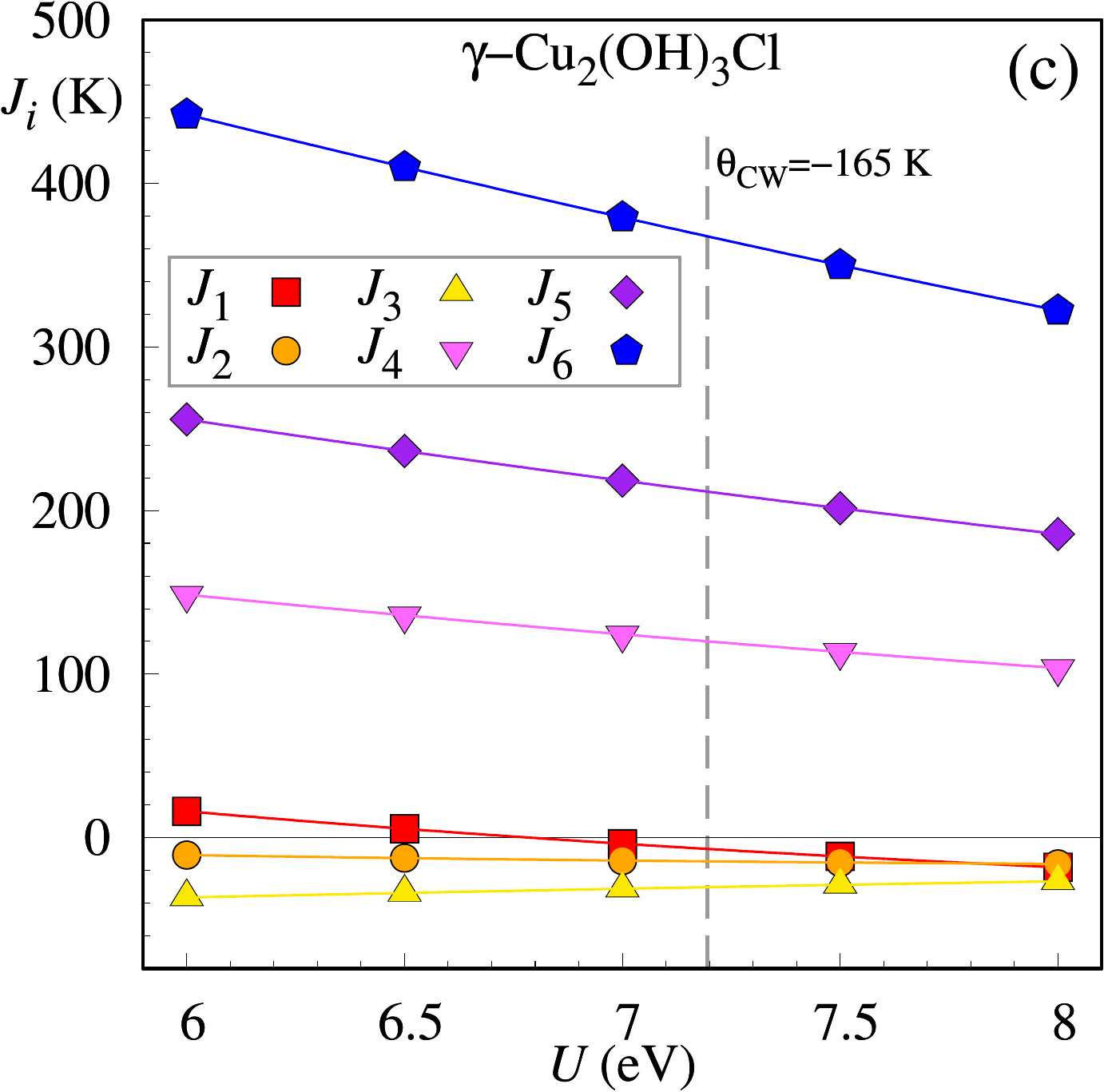}
\caption{(a) Crystal structure of clinoatacamite~\cite{Malcherek2009} viewed along the $b$ axis [Cu(1): teal, Cu(2): light blue, Cu(3): dark blue, Cl: green, O: red, H: white]. (b) Motif of NN Cu-Cu exchange interactions $J_1$--$J_6$ and (c) their strength derived by means of DFT and GGA+$U$, plotted as function of the on-site Coulomb repulsion $U$. At $U = 7.19$\,eV, the mean-field Curie-Weiss temperature agrees with the experimental value $\Theta_\mathrm{CW} = -165(1)$\,K with $J_1 = -6.8$\,K, $J_2 = -14.5$\,K, $J_3 = -30.3$\,K, $J_4 = 120.1$\,K, $J_5 = 211.6$\,K and $J_6 = 367.7$\,K (details in Ref.~\cite{SI}). In panels (a) and (b), the unit cells are indicated.}
\label{fig:Fig1}
\end{figure*}

Distortions of a spin lattice impair the equivalence of magnetic sites. At first sight, this should lift the underconstrained nature of a kagome antiferromagnet and select distinct ordering modes. However, the manifold of ordering modes may decompose into a system of modes pertaining to different symmetries. Their competition can reestablish magnetic frustration at a higher level, resulting in complex magnetic behavior. In the present study, we address the unusual magnetic properties of one such system, clinoatacamite Cu$_2$Cl(OH)$_3$, a distorted kagome material and a longstanding puzzle~\cite{Zheng2005ClinoPRL} in frustrated quantum magnetism. As parent compound of the paratacamite family Zn$_x$Cu$_{4-x}$Cl$_2$(OH)$_6$, clinoatacamite ($x=0$) is closely related to herbertsmithite ($x=1$), but its kagome motif of Cu sites is embedded into a low-symmetry crystal structure with different symmetries for the kagome sites. Further, unlike the QSL candidate herbertsmithite, clinoatacamite undergoes a magnetic transition at 18.1\,K~\cite{Zheng2005ClinoPRL}. 

Although actively discussed in the context of kagome physics in the past~\cite{Lee2007,Kim2008,Wills2008,Wills2009,Maegawa2010,Morodomi2010,Morodomi2012}, the ground states of clinoatacamite have remained inconclusive~\cite{comm:inconclusive}, in part because of the lack of single-crystal studies. In this Letter, employing density-functional theory (DFT) we will first demonstrate that clinoatacamite can be understood as a non-uniform kagome antiferromagnet with weak ferromagnetic (FM) couplings to the interlayer Cu sites. Hence, it clearly distinguishes itself from other kagome materials in the atacamite family~\cite{Puphal2018} leading to a weak network of kagome planes and interlayer sites. We will further demonstrate using thermodynamic as well as microscopic measurement techniques and single-crystalline samples that clinoatacamite exhibits a complex zero-field sequence of magnetic states below $T_\mathrm{I} = 18.1$\,K. Zero-field muon spin rotation/relaxation ($\mu$SR) and neutron diffraction reveal different types of long-range order which separate the temperature range below $T_\mathrm{I}$ into three regions I-III. We further find that the multicritical nature of the distorted kagome compound clinoatacamite manifests itself in the $\mathbf{q}_\mathrm{m} = (0,0,0)$ state at lowest temperatures. It consists of simultaneous odd-even AFM ordering modes implying a metamagnetic texture~\cite{Sokolov2019}. With this, clinoatacamite serves as an example of a frustrated spin-$1/2$ antiferromagnet with complex magnetic behavior in zero field. To account for this, we combine complementary measurement techniques, microscopic details as well as phenomenology for multicritical systems. 

Clinoatacamite crystallizes in a monoclinic structure of space group $P2_1/n$ (No.\,14)~\cite{Grice1996} and with lattice parameters $a=6.1675(7)$\,\AA, $b=6.8327(8)$\,\AA, $c=9.1517(4)$\,\AA~and $\beta = 99.492(2)^{\circ}$~\cite{Malcherek2009} [Fig.~\ref{fig:Fig1}(a)]. Three inequivalent copper sites Cu(1)--Cu(3) are present. As depicted in Fig.~\ref{fig:Fig1}(a), Cu(1)(OH)$_6$ and Cu(2)(OH)$_4$Cl$_2$ octahedra alternate along the $a$ axis. Cu(3)(OH)$_4$Cl$_2$ octahedra form chains along the $b$ axis. The Cu(2) and Cu(3) sites form a slightly distorted kagome motif, spanned by the $[010]$ and $[10\overline{1}]$ crystal directions. Cu(1) sites are interlayer sites. There are three inequivalent NN Cu-Cu exchange paths ($J_4$-$J_6$) within the kagome layers and three inequivalent NN Cu-Cu exchange paths ($J_1$-$J_3$) between kagome and interlayer sites [Fig.~\ref{fig:Fig1}(b)].

We show that the kagome motif is formed by the dominant exchange couplings by employing DFT with full potential local orbital basis~\cite{Koepernik1999} and generalized gradient approximation (GGA) functional~\cite{Perdew1996}. Electronic correlations on the Cu$^{2+}$ ions were accounted for by GGA+$U$~\cite{Liechtenstein1995}. As the strength of the exchange couplings is well known to sensitively depend on Cu--O--Cu bond angles, we have first studied the crystal structure of clinoatacamite using single-crystal neutron diffraction. We collected the data using the Four-Circle Diffractometer E5 at the BER II reactor of the Helmholtz-Zentrum Berlin f\"ur Materialien und Energie (HZB), Germany~\cite{SI,Xtal3.4,Sears1995,Sears1992_absorption,Zachariasen1967}. Subsequently, we used the obtained positional parameters as input for the DFT-based analysis together with lattice parameters from Ref.~\cite{Malcherek2009}. Fig.~\ref{fig:Fig1}(c) visualizes the $U$ dependence of $J_1$-$J_6$. Next-nearest neighbor exchanges $J_7$-$J_{29}$ were found to be $\left|J_i\right| < 10$\,K~\cite{SI}. We calculated the $J_i$ at $U = 7.19$\,eV, where the mean-field Curie-Weiss temperature agrees with the experimental value $\Theta_\textrm{CW}=-165(1)$\,K extracted from a fit of the inverse magnetic susceptibility~\cite{SI}. 

The dominant exchange couplings, $J_4 = 120.1$\,K, $J_5 = 211.6$\,K and $J_6 = 367.7$\,K form the non-uniform AFM kagome layers. Here, $J_6$ couples units of three Cu sites into trimers, similar to the motif deduced in Ref.~\cite{Kim2008} for in-plane exchange only~\cite{comm:trimermodel,Harris2013}. The couplings $J_1 = -6.8$\,K, $J_2 = -14.5$\,K and $J_3= -30.3$\,K are FM, as suggested in Ref.~\cite{Khatami2012}. These couplings between kagome sites and interlayer sites $\left|J_i\right| \lesssim 0.082\,J_6$ are somewhat stronger than the interlayer couplings in herbertsmithite~\cite{Jeschke2013} where nonmagnetic Zn occupies the interlayer sites. In clinoatacamite, this leads to a weak ferromagnetic network between kagome and interlayer sites which stabilizes long-range order. Still, the FM couplings to the interlayer Cu sites are significantly smaller than in the mineral barlowite Cu$_4$(OH)$_6$FBr with uniform AFM kagome planes~\cite{Jeschke2015}. In a classical Monte Carlo approach for the ordered magnetic ground state resulting from the Hamiltonian of clinoatacamite we find that the interlayer Cu(1) spins adopt a spiral order reflecting the weak coupling to the kagome spins with up-up-down order ($T_\mathrm{N}^\mathrm{cMC} = 13$\,K; details in Ref.~\cite{SI}).

\begin{figure}[t!]
\centering
\includegraphics[width=\linewidth]{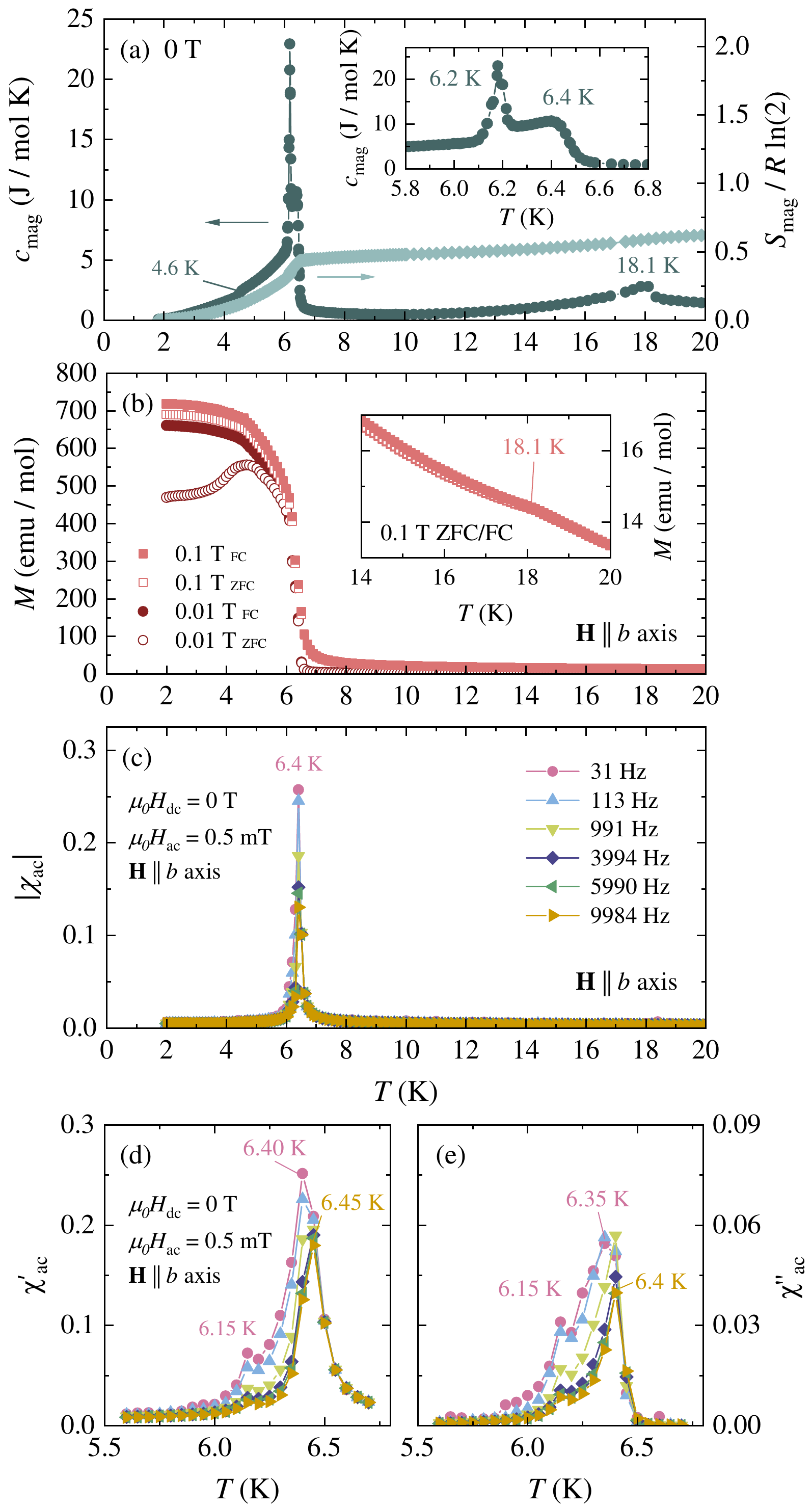}
\caption{(a) Magnetic specific heat $c_\mathrm{mag}$ and entropy $S_\mathrm{mag}$ in zero field and as function of temperature. The inset enlarges the region of the maxima at 6.2\,K and 6.4\,K. (b) Temperature dependence of the magnetization in weak dc magnetic fields (0.01\,T, 0.1\,T) for FC and ZFC samples ($\mathbf{H} \parallel b$ axis). The inset shows the subtle kink at $18.1$\,K in 0.1\,T. (c) Temperature dependence of the magnitude of the ac magnetic susceptibility $\left|\chi_\mathrm{ac}\right| = \sqrt{\chi^{\prime 2}_\mathrm{ac} + \chi^{\prime\prime 2}_\mathrm{ac}}$ with $\mu_0H_\mathrm{dc} = 0$\,T and $\mu_0H_\mathrm{ac} = 0.5$\,mT ($\mathbf{H} \parallel b$ axis) for various frequencies up to 9984\,Hz. (d)--(e) In-phase and out-of-phase components of the ac susceptibility, $\chi^{\prime}_\mathrm{ac}$ and $\chi^{\prime\prime}_\mathrm{ac}$, in the region of the 6.2\,K and 6.4\,K anomalies.}
\label{fig:Fig2}
\end{figure}

Experimentally, the use of natural single-crystalline samples (details in Ref.~\cite{SI}) allows us to resolve a complex sequence of magnetic phases/regions in zero and weak magnetic fields ($\mathbf{H} \parallel b$ axis). As shown in Fig.~\ref{fig:Fig2}(a), the magnetic specific heat $c_\mathrm{mag}$~\cite{SI} features an anomaly at 18.1\,K in zero field, in full agreement with earlier results~\cite{Zheng2005Clino,Zheng2005ClinoPRL}. Additionally, we observe two transition anomalies at 6.2\,K and 6.4\,K [inset Fig.~\ref{fig:Fig2}(a)], not resolved as distinct maxima in previous polycrystal studies~\cite{Zheng2005Clino}. At 4.6\,K, we detect a weak kink in the specific heat not reported before. The magnetic entropy $S_\mathrm{mag}$ at 18.1\,K, obtained via an integration of $c_\mathrm{mag}/T$~\cite{SI}, is only $\sim 1/3$ of $2\,R\,\ln(2)$, the entropy expected for two spin-$1/2$ per formula unit. The entropy change at 18.1\,K is very small. The very sharp and intense 6.2\,K-anomaly resembles that of a first-order phase transition. 

Magnetometry in weak dc magnetic fields $\mathbf{H} \parallel b$ axis reveals different behavior below 6.2\,K for field-cooled (FC) and zero-field cooled (ZFC) samples up to $\sim 0.1$\,T [Fig.~\ref{fig:Fig2}(b)] in accordance with earlier results for polycrystals~\cite{Zheng2005Clino}. At 4.7\,K, slightly shifted from the temperature of the weak anomaly in the specific heat [Fig.~\ref{fig:Fig2}(a)], the $M(T)$-data feature a kink/maximum at 0.01\,T (FC/ZFC). The magnetization anomaly at 18.1\,K is very weak and is visible on a reduced scale only [inset Fig.~\ref{fig:Fig2}(b)]. The ac magnetic susceptibility $\chi_\mathrm{ac}$ [Fig.~\ref{fig:Fig2}(c)] features an intense anomaly at 6.4\,K for all frequencies studied ($\mu_0H_\mathrm{dc} = 0$\,T, $\mu_0H_\mathrm{ac} = 0.5$\,mT). A frequency-dependent behavior of $\left|\chi_\mathrm{ac}\right|$ sets in at 6.4\,K and persists down to $\sim 4.5$\,K, where a weak kink-like anomaly is present, not visible on the scale of Fig.~\ref{fig:Fig2}(c). No anomaly can be resolved at 18.1\,K. In Figs.~\ref{fig:Fig2}(d)--(e), in-phase and out-of-phase components of the ac susceptibility, $\chi^{\prime}_\mathrm{ac}$ and $\chi^{\prime\prime}_\mathrm{ac}$, are plotted in the temperature region of the 6.2\,K and 6.4\,K transitions where the frequency dependence is most pronounced. Our observations are at variance with the behavior presented in the past for synthetic, polycrystalline clinoatacamite, where $\chi^{\prime}_\mathrm{ac}$ was found to be clearly frequency-dependent and $\chi^{\prime\prime}_\mathrm{ac}$ to be non-zero even below 5\,K~\cite{Zheng2005Clino}. Based on those earlier results, a spin-glass-like ground state was stated. Our results rule out spin-glassiness at low temperatures.

\begin{figure}[t!]
\centering
\includegraphics[width=\linewidth]{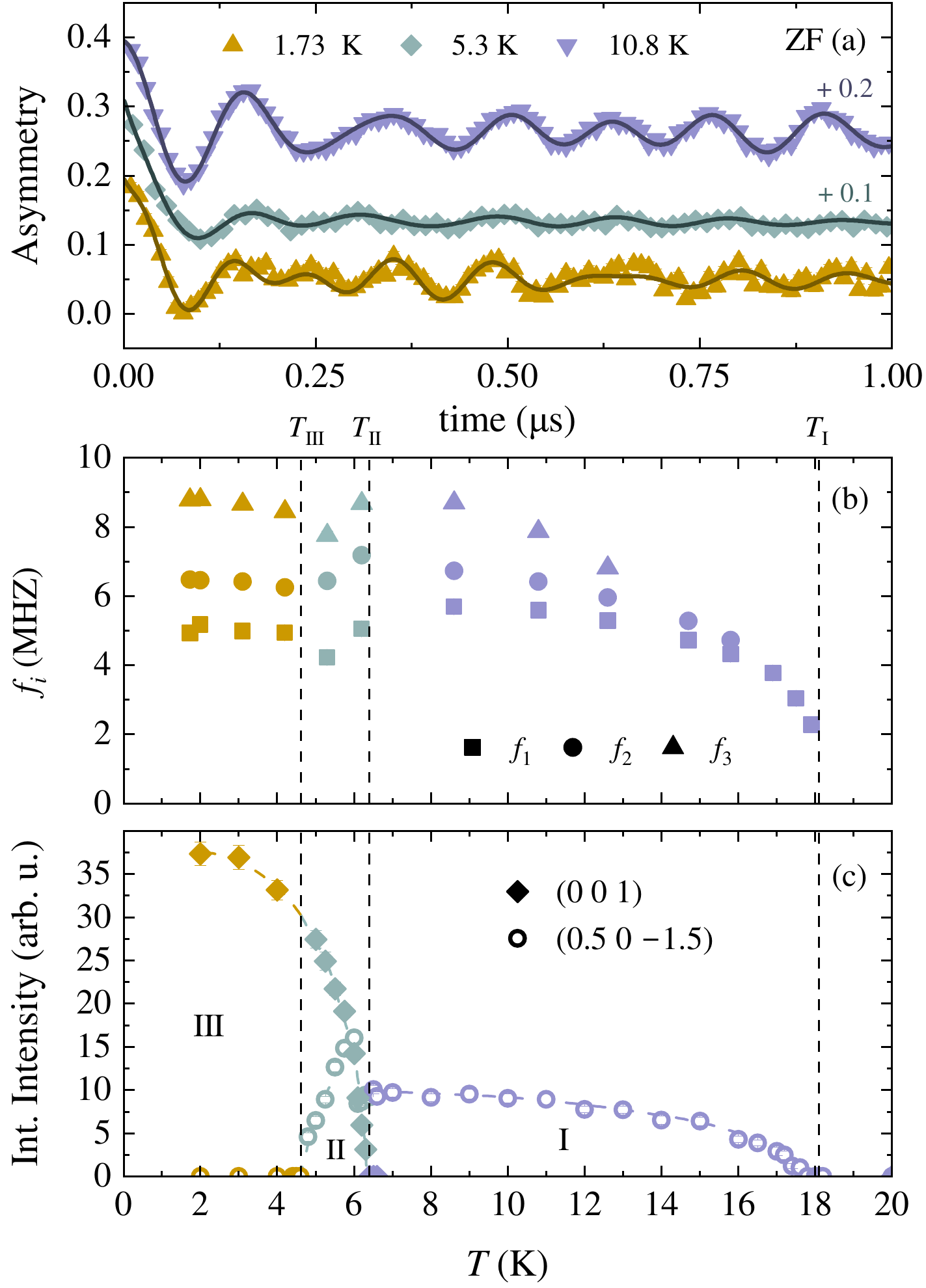}
\includegraphics[width=0.97\linewidth]{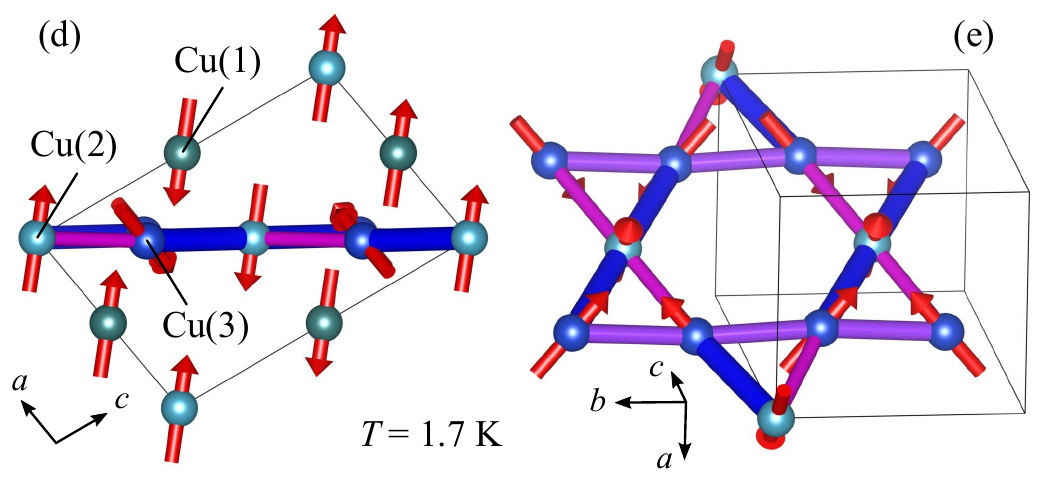}
\caption{(a) Zero-field $\mu$SR asymmetry of single-crystalline clinoatacamite as function of time measured at 1.73\,K, 5.3\,K and 10.8\,K. The data at 5.3\,K and 10.8 K are shifted vertically for clarity. Solid curves are fits to the data. (b) Temperature dependence of the muon spin precession frequencies $f_i$ determined from the fits; for details see Ref.~\cite{SI}. (c) Temperature dependence of the integrated intensities of the purely magnetic $(0\,0\,1)$ and $(0.5\,0\,{-1.5})$ reflections from neutron diffraction in zero field. (d)--(e) Magnetic structure at 1.7\,K [$R_\mathrm{M} = 0.0545$ (in $F$)] together with the kagome motif viewed along the $b$ axis and (e) displaying a single distorted kagome unit formed by Cu(2)/Cu(3) sites.}
\label{fig:Fig3}
\end{figure}

We complement the view on the zero-field sequence of phases/regions by carrying out single-crystal $\mu$SR as well as neutron diffraction experiments. Zero-field (ZF) and weak transverse field (wTF) $\mu$SR measurements were carried out using the General Purpose Surface-Muon Instrument~\cite{GPSpaper} of the Swiss Muon Source at the Paul Scherrer Institut, Switzerland. $\mu$SR clearly senses long-range magnetic order below 18.1\,K. In wTF (30\,G) at temperatures $T > 18.1$\,K, in the paramagnetic phase, the whole asymmetry oscillates as the muon spins precess with the frequency of the external field~\cite{SI}. Below 18.1\,K the asymmetry in wTF is lost indicating that long-range order is present. Exemplary ZF asymmetry spectra as function of time taken at 1.73\,K, 5.3\,K and 10.8\,K are presented in Fig.~\ref{fig:Fig3}(a). They are typical for the temperature regions below 4.6\,K, above 6.4\,K and the region between these two thermodynamic transitions, and they indicate different types of long-range magnetic order. While above 6.4\,K and below 4.6\,K the experimental data can be well fitted to three distinct precession frequencies due to internal fields [Fig.~\ref{fig:Fig3}(a)--(b)], the situation in the intermediate range is more complex with indications for enhanced spin dynamics similar to the findings on polycrystalline material reported in Ref.~\cite{Zheng2005ClinoPRL}. An extensive analysis of our $\mu$SR data with detailed discussion will be given elsewhere.

In agreement with previous results for deuterated clinoatacamite powder~\cite{Lee2007,Kim2008,Wills2008}, our single-crystal neutron diffraction study reveals magnetic reflections below 6.4\,K which are compatible with the magnetic propagation vector $\mathbf{q}_\mathrm{m} = (0, 0, 0)$~\cite{SI}. Neutron diffraction experiments~\cite{ILL_5-41-1083,ILL_EASY-1612,ILL-5-41-1349} were carried out using the Four-Circle Diffractometer D10 of the Institut Laue-Langevin (ILL), France, as well as the Flat-Cone Diffractometer E2~\cite{E2paper} at HZB, Germany. As displayed in Fig.~\ref{fig:Fig3}(c), the integrated intensity of the purely magnetic $(0\,0\,1)$ reflection decreases upon an increase of temperature and vanishes at 6.4(1)\,K, signaling the suppression of the $\mathbf{q}_\mathrm{m} = \mathbf{0}$ state. 

In the temperature region directly below $T_\mathrm{I} = 18.1$\,K, we observe magnetic reflections which can be described by the magnetic propagation vector $\mathbf{q}_\mathrm{m} = (-0.5, 0, 0.5)$ which also was observed at 7\,K during a recent neutron diffraction study on deuterated powder~\cite{Zheng2024}. In Fig.~\ref{fig:Fig3}(c) we present the temperature dependence of the integrated intensity of the magnetic $(0.5\,0\,{-1.5})$ reflection. It goes through an intensity maximum at 6.0\,K and persists down to 4.6(1)\,K. As a result, the magnetic states with propagation vectors $\mathbf{q}_\mathrm{m} = (0, 0, 0)$ and $\mathbf{q}_\mathrm{m} = (-0.5, 0, 0.5)$ coexist between $T_\mathrm{III} = 4.6$\,K and $T_\mathrm{II} = 6.4$\,K, which is where the $\mu$SR fits reflected a more complex behavior. We label the regions~I (6.4\,K--18.1\,K), II (4.6\,K--6.4\,K) and III (below 4.6\,K) as indicated in Fig.~\ref{fig:Fig3}(b)--(c). Our findings fully resolve the apparent inconsistency between neutron diffraction and $\mu$SR data which has persisted for decades~\cite{comm:inconclusive}. While the determination of the magnetic structure in region I will be given elsewhere, we reveal that the competition of ordering modes already manifests itself in the $\mathbf{q}_\mathrm{m} = (0, 0, 0)$ order in region~III. Our findings further rule out the scenario of clinoatacamite as valence bond solid from Ref.~\cite{Lee2007}.

Key to the determination of the magnetic structure in region III was to measure the intensity of magnetic reflections, weak ones in particular, with good accuracy. This was limited in earlier powder studies~\cite{Lee2007,Kim2008,Wills2008}, whereas single-crystal neutron diffraction allows to record the integrated intensities of individual Bragg peaks. Based on a corresponding experiment on D10~\cite{ILL_5-41-1083}, we have carried out a refinement of the magnetic structure employing representation analysis~\cite{Bertaut1968,Fullprof,Brown1995} (details in Ref.~\cite{SI}). We found that the data are best described [$R_\mathrm{M} = 0.0545$ (in $F$)] using different irreducible representations for Cu(1)/Cu(2) on Wyckoff sites $2d$/$2a$ and Cu(3) on $4e$. The resulting magnetic structure is displayed in Figs.~\ref{fig:Fig3}(d)--(e). The magnetic moments of Cu(1)/Cu(2)~\cite{comm:Cu1Cu2} lie in the $ac$ plane with components of $\mu_\mathrm{ord, Cu(1)}(x) = \mu_\mathrm{ord, Cu(2)}(x) = 0.30(2)\,\mu_\mathrm{B}$ and $\mu_\mathrm{ord, Cu(1)}(z) = \mu_\mathrm{ord, Cu(2)}(z) = 0.29(3)\,\mu_\mathrm{B}$, resulting in a total moment of $\mu_\mathrm{ord, Cu(1)} = \mu_\mathrm{ord, Cu(2)} = 0.38(2)\,\mu_\mathrm{B}$. The magnetic moments of Cu(3) lie in the $ab$ plane with $\mu_\mathrm{ord, Cu(3)}(x) = 0.29(2)\,\mu_\mathrm{B}$ and $\mu_\mathrm{ord, Cu(3)}(y) = 0.25(2)\,\mu_\mathrm{B}$ resulting in a total moment of $\mu_\mathrm{ord, Cu(3)} = 0.38(2)\,\mu_\mathrm{B}$. The refinement yields collinear Cu(1)/Cu(2) moments, almost normal to the kagome plane, produced by an even-parity ordering mode (inversion-symmetric). The non-collinear Cu(3) moments are produced by an odd-parity ordering mode (broken inversion symmetry).

As the frustration of clinoatacamite is strong, below 18.1\,K, these symmetry-different modes are expected to be close to multicriticality. This can be understood by considering the deformation of the underconstrained undistorted kagome antiferromagnet which supports many degenerate modes. Some of these modes must be connected with the distinct selected modes dominating magnetic ordering along the distortion path. From phenomenological Landau-theory for multicritical systems, one of the two competing modes should yield long-range order by a conventional continuous phase transition upon lowering temperature from the paramagnetic state. Eventually, a transition to a state of coexisting modes, or a domain state involving other modes could be seen in a bi- or tetracritical phase diagram~\cite{ChaikinLubensky1995}.

From phenomenology of the multicritical system of modes, the behavior of clinoatacamite can be rationalized. Owing to the even and odd parity of the ordering modes, the magnetic free energy contains so-called {\em Lifshitz-type} invariants (LtIs), bilinear terms with one gradient~\cite{SI}. These coupling terms allow the co-existence of symmetry-different modes in a common texture, a spatially modulated state in the order-parameter space as a homogeneous long-range order of one mode is destabilized against a spatially varying admixture of the other mode. The mechanism is similar to the rotation of the magnetization vector in a chiral helimagnet~\cite{Dzyaloshinskii1964}. Such modulated mixtures of symmetry-different modes have been observed so far only in non-centrosymmetric magnetic crystals: as {\em metamagnetic texture}, a field-driven mixture of an AFM configuration with a FM state~\cite{Sokolov2019}, while in a highly frustrated magnetic system similar textures have been invoked to explain the observation of quasi-static lumps of non-collinear spin-structures in absence of any long-range order~\cite{Schweika2022}. The spin configuration below $T_\mathrm{II} = 6.4$\,K is a mixture of competing symmetry-different modes. This confirms the multicritical nature of magnetic ordering in clinoatacamite. Hence, the transition at $T_\mathrm{I} = 18.1$\,K could signal the formation of a textured state dominated by a primary mode frustrated by the closeness of another mode. The formation of such a partial order can explain the weak thermodynamic signature of this transition from the paramagnetic state. Consequently, the complex transition at $T_\mathrm{II}$ appears to be triggered when the second ordering mode becomes thermodynamically stable upon lowering temperature. This process should reorganize the textured high-temperature state into a long-range ordered phase, which may explain the observation of a sequence of narrow intermediate or mesophase states at this transition.

In conclusion, clinoatacamite exemplifies a geometrically frustrated antiferromagnet where the existence of an underconstrained manifold of modes allows for a mechanism where competing modes can coexist and create textured magnetic states of mixed symmetry. The even/odd parity of the ordering modes is instrumental for such states. Such systems become exceedingly complex because of the multicriticality and the higher-level, i.e. non-local, frustration through the chiral Lifshitz invariants. In addition, in a low-symmetry crystal like clinoatacamite, many anisotropic coupling terms will influence its magnetic phase diagram.

\begin{acknowledgments}
\textit{Acknowledgements}---We thank the Helmholtz-Zentrum Berlin f{\"u}r Materialien und Energie as well as the Institut Laue-Langevin for allocation of neutron radiation beamtime and thankfully acknowledge the financial support from HZB and ILL. This work is further based on experiments performed at the Swiss Muon Source S\textmu{S}, Paul Scherrer Institute, Villigen, Switzerland. We thankfully acknowledge technical support from B.~Veltel using the Physics Laboratory of MLZ. We thank T.~Lampe and A. Kriele for their support during the sample preparation at the TU Braunschweig and at the Materials Science Lab operated by Helmholtz-Zentrum Hereon at the Heinz Maier Leibnitz Zentrum (MLZ), Germany. We further thank J.~L.~Winter, K.~Koschel and J.~Willwater for their support in the early stages of this project. This work has partially been supported by the DFG under Contract No.~SU229/9-2, through SFB 1143 (Project No. 247310070) and Project No. 390858490 (W\"urzburg-Dresden Cluster of Excellence on Complexity and Topology in Quantum Matter--ct.qmat, EXC 2147). H.~O.~J. acknowledges support through JSPS KAKENHI Grants No.~24H01668 and No.~25K0846007. F.~B.~acknowledges support from the Agence Nationale de la Recherche (ANR), project ANR-25-CE30-2010-01. U.~K.~R.~thanks U.~Nitzsche for assistance with computational resources. Crystal and magnetic structures were drawn using VESTA 3~\cite{VESTA}.
\end{acknowledgments}


\begin{thebibliography}{999}
\bibitem{Balents2010} L. Balents, Spin liquids in frustrated magnets, \href{https://doi.org/10.1038/nature08917}{Nature \textbf{464}, 199 (2010)}.
\bibitem{Broholm2020} C. Broholm, R. J. Cava, S. A. Kivelson, D. G. Nocera, M. R. Norman, T. Senthil, Quantum Spin Liquids, \href{https://doi.org/10.1126/science.aay0668}{Science \textbf{367}, eaay0668 (2020)}.
\bibitem{Mendels2007} P. Mendels, F. Bert, M. A. de Vries, A. Olariu, A. Harrison, F. Duc, J. C. Trombe, J. S. Lord, A. Amato, and C. Baines, Quantum Magnetism in the Paratacamite Family: Towards an Ideal Kagom\'e Lattice, \href{https://doi.org/10.1103/PhysRevLett.98.077204}{Phys. Rev. Lett. \textbf{98}, 077204 (2007)}.
\bibitem{Norman2016} M. R. Norman, \textit{Colloquium}: Herbertsmithite and the search for the quantum spin liquid, \href{https://doi.org/10.1103/RevModPhys.88.041002}{Rev. Mod. Phys. \textbf{88}, 041002 (2016)}.
\bibitem{Khuntia2020} P. Khuntia, M. Vel\'azquez, Q. Barth\'elemy, F. Bert, E. Kermarrec, A. Legros, B. Bernu, L. Messio, A. Zorko, and P. Mendels, Gapless ground state in the archetypal quantum kagome antiferromagnet ZnCu$_3$(OH)$_6$Cl$_2$, \href{https://doi.org/10.1038/s41567-020-0792-1}{Nat. Phys. \textbf{16}, 469 (2020)}.
\bibitem{Kermarrec2011} E. Kermarrec, P. Mendels, F. Bert, R. H. Colman, A. S. Wills, P. Strobel, P. Bonville, A. Hillier, and A. Amato, Spin-liquid ground state in the frustrated kagome antiferromagnet MgCu$_3$(OH)$_6$Cl$_2$, \href{https://doi.org/10.1103/PhysRevB.84.100401}{Phys. Rev. B \textbf{84}, 100401(R) (2011)}.
\bibitem{Boldrin2015} D. Boldrin, B. F\r{a}k, M. Enderle, S. Bieri, J. Ollivier, S. Rols, P. Manuel, and A. S. Wills, Haydeeite: A spin-$1/2$ kagome ferromagnet, \href{https://doi.org/10.1103/PhysRevB.91.220408}{Phys. Rev. B \textbf{91}, 220408(R) (2015)}.
\bibitem{Puphal2018} P. Puphal, K. M. Zoch, J. D\'esor, M. Bolte, and C. Krellner, Kagom\'e quantum spin systems in the atacamite family, \href{https://doi.org/10.1103/PhysRevMaterials.2.063402}{Phys. Rev. Mater. \textbf{2}, 063402 (2018)}.
\bibitem{Okuma2019} R. Okuma, D. Nakamura, T. Okubo, A. Miyake, A. Matsuo, K. Kindo, M. Tokunaga, N. Kawashima, S. Takeyama, and Z. Hiroi, A series of magnon crystals appearing under ultrahigh magnetic fields in a kagomé antiferromagnet, \href{https://doi.org/10.1038/s41467-019-09063-7}{Nat. Commun. \textbf{10}, 1229 (2019)}.
\bibitem{Henderson2019} A. Henderson, L. Dong, S. Biswas, H. I. Revell, Y. Xin, R. Valenti, J. A. Schlueter, and T. Siegrist, Order-disorder transition in the $S = 1/2$ kagome antiferromagnets claringbullite and barlowite, \href{https://doi.org/10.1039/C9CC04930D}{Chem. Commun. \textbf{55}, 11587 (2019)}.
\bibitem{Akazawa2020} M. Akazawa, M. Shimozawa, S. Kittaka, T. Sakakibara, R. Okuma, Z. Hiroi, H.-Y. Lee, N. Kawashima, J. H. Han, and M. Yamashita, Thermal Hall Effects of Spins and Phonons in Kagome Antiferromagnet Cd-Kapellasite \href{https://doi.org/10.1103/PhysRevX.10.041059}{Phys. Rev. X \textbf{10}, 041059 (2020)}.
\bibitem{Tustain2021} K. Tustain, E. E. McCabe, A. M. Arevalo-Lopez, A. S. Gibbs, S. P. Thompson, C. A. Murray, C. Ritter and L. Clark, Disorder-Induced Structural Complexity in the Barlowite Family of $S = 1/2$ Kagomé Magnets, \href{https://doi.org/10.1021/acs.chemmater.1c03247}{Chem. Mater. \textbf{33}, 24, 9638 (2021)}.
\bibitem{Chatterjee2023} D. Chatterjee, P. Puphal, Q. Barthélemy, J. Willwater, S. S\"ullow, C. Baines, S. Petit, E. Ressouche, J. Ollivier, K. M. Zoch, C. Krellner, M. Parzer, A. Riss, F. Garmroudi, A. Pustogow, P. Mendels, E. Kermarrec, and F. Bert, From spin liquid to magnetic ordering in the anisotropic kagome Y-kapellasite Y$_3$Cu$_9$(OH)$_{19}$Cl$_8$: A single-crystal study, \href{https://doi.org/10.1103/PhysRevB.107.125156}{Phys. Rev. B \textbf{107}, 125156 (2023)}.
\bibitem{Wang2023} J. Wang, M. Spitaler, Y.-S. Su, K. M. Zoch, C. Krellner, P. Puphal, S. E. Brown, and A. Pustogow, Controlled Frustration Release on the Kagome Lattice by Uniaxial-Strain Tuning, \href{https://doi.org/10.1103/PhysRevLett.131.256501}{Phys. Rev. Lett. \textbf{131}, 256501 (2023)}.

\bibitem{Matan2010} K. Matan, T. Ono, Y. Fukumoto, T. J. Sato, J. Yamaura, M. Yano, K. Morita, and H. Tanaka, Pinwheel valence-bond solid and triplet excitations in the two-dimensional deformed kagome lattice, \href{https://doi.org/10.1038/nphys1761}{Nat. Phys. \textbf{6}, 865 (2010)}.
\bibitem{Grbic2013} M. S. Grbić, S. Kr\"amer, C. Berthier, F. Trousselet, O. Cépas, H. Tanaka, and M. Horvatić, Microscopic Properties of the Pinwheel Kagome Compound Rb$_2$Cu$_3$SnF$_{12}$, \href{https://doi.org/10.1103/PhysRevLett.110.247203}{Phys. Rev. Lett. \textbf{110}, 247203 (2013).}
\bibitem{Hering2022} M. Hering, F. Ferrari, A. Razpopov, I. I. Mazin, R. Valentí, H. O. Jeschke, and J. Reuther, Phase diagram of a distorted kagome antiferromagnet and application to Y-kapellasite, \href{https://doi.org/10.1038/s41524-021-00689-0}{npj Comput Mater 8, \textbf{10} (2022)}.
\bibitem{Janson2016} O. Janson, S. Furukawa, T. Momoi, P. Sindzingre, J. Richter, and K. Held, Magnetic Behavior of Volborthite Determined by Coupled Trimers Rather than Frustrated Chains, \href{https://doi.org/10.1103/PhysRevLett.117.037206}{Phys. Rev. Lett. \textbf{117}, 037206 (2016)}.
\bibitem{Matan2019} K. Matan, T. Ono, G. Gitgeatpong, K. de Roos, P. Miao, S. Torii, T. Kamiyama, A. Miyata, A. Matsuo, K. Kindo, S. Takeyama, Y. Nambu, P. Piyawongwatthana, T. J. Sato, and H. Tanaka, Magnetic structure and high-field magnetization of the distorted kagome lattice antiferromagnet Cs$_2$Cu$_3$SnF$_{12}$, \href{https://doi.org/10.1103/PhysRevB.99.224404}{Phys. Rev. B \textbf{99}, 224404 (2019)}.

\bibitem{Zheng2005ClinoPRL} X. G. Zheng, H. Kubozono, K. Nishiyama, W. Higemoto, T. Kawae, A. Koda, and C. N. Xu, Coexistence of Long-Range Order and Spin Fluctuation in Geometrically Frustrated Clinoatacamite Cu$_{2}$Cl(OH)$_{3}$, \href{https://doi.org/10.1103/PhysRevLett.95.057201}{Phys. Rev. Lett. \textbf{95}, 057201 (2005)}.
\bibitem{Lee2007} S.-H. Lee, H. Kikuchi, Y. Qiu, B. Lake, Q. Huang, K. Habicht, and K. Kiefer, Quantum-spin-liquid states in the two-dimensional kagome antiferromagnets Zn$_x$Cu$_{4-x}$(OD)$_6$Cl$_2$, \href{https://doi.org/10.1038/nmat1986}{Nat. Mater. \textbf{6}, 853 (2007)}.

\bibitem{Kim2008} J.-H. Kim, S. Ji, S.-H. Lee, B. Lake, T. Yildirim, H. Nojiri, H. Kikuchi, K. Habicht, Y. Qiu, and K. Kiefer, External Magnetic Field Effects on a Distorted Kagome Antiferromagnet, \href{https://doi.org/10.1103/PhysRevLett.101.107201}{Phys. Rev. Lett. \textbf{101}, 107201 (2008)}.
\bibitem{Wills2008} A. S. Wills and J.-Y. Henry, On the crystal and magnetic ordering structures of clinoatacamite, $\gamma$-Cu$_{2}$Cl(OH)$_{3}$, a proposed valence bond solid, \href{https://doi.org/10.1088/0953-8984/20/47/472206}{J. Phys.: Condens. Matter \textbf{20}, 472206 (2008)}.
\bibitem{Wills2009} A. S. Wills, T. G. Perring, S. Raymond, B. F\r{a}k, J.-Y. Henry, and M. Telling, Inelastic neutron scattering studies of the quantum frustrated magnet clinoatacamite, $\gamma$-Cu$_2$(OD)$_3$Cl, a proposed valence bond solid (VBS), \href{https://doi.org/10.1088/1742-6596/145/1/012056}{J. Phys.: Conf. Ser. \textbf{145}, 012056 (2009)}.
\bibitem{Maegawa2010} S. Maegawa, A. Oyamada, and S. Sato, Novel Frustrated Behavior in Quantum Heisenberg Antiferromagnets on the Pyrochlore Lattice: NMR Studies of R$_2$(OH)$_3$Cl (R = Cu and Ni), \href{https://doi.org/10.1143/JPSJ.79.011002}{J. Phys. Soc. Jpn. \textbf{79}, 011002 (2010)}.
\bibitem{Morodomi2010} H. Morodomi, K. Ienaga, Y. Inagaki, T. Kawae, M. Hagiwara, X. G. Zheng, Specific heat study of geometrically frustrated magnet clinoatacamite Cu$_{2}$Cl(OH)$_{3}$, \href{https://doi.org/10.1088/1742-6596/200/3/032047}{J. Phys.: Conf. Ser. \textbf{200}, 032047 (2010)}.
\bibitem{Morodomi2012} H. Morodomi, K. Ienaga, Y. Inagaki, T. Kawae, M. Hagiwara, X. G. Zheng, Magnetic Field dependence of specific heat in Clinoatacamite Cu$_{2}$Cl(OH)$_{3}$, \href{https://doi.org/10.1088/1742-6596/400/3/032058}{J. Phys.: Conf. Ser. \textbf{400}, 032058 (2012)}. 
\bibitem{comm:inconclusive} In the past, magnetic studies on clinoatacamite were carried out using synthesized (in part deuterated) as well as natural polycrystals and powder~\cite{Wills2004,Zheng2005Clino,Zheng2005ClinoPRL,Lee2007,Kim2008,Wills2008,Wills2009,Maegawa2010,Morodomi2010,Morodomi2012}, however, the results yielded an inconsistent picture. It was found that clinoatacamite undergoes a magnetic transition at $18.1$\,K, signaled by weak anomalies of the magnetic susceptibility and the heat capacity as well as by the onset of oscillations of the $\mu$SR asymmetry~\cite{Zheng2005ClinoPRL,Zheng2005Clino}. An additional transition was reported to occur at 6.2\,K~\cite{Zheng2005ClinoPRL}. Zero-field $\mu$SR measurements in this lower-temperature region revealed oscillations of the asymmetry only at lowest temperatures $\sim 2$\,K~\cite{Zheng2005ClinoPRL}. Powder neutron scattering of deuterated material supported long-range magnetic order with a propagation vector $\mathbf{q}_\mathrm{m}=(0,0,0)$ below $6.7$\,K~\cite{Lee2007,Kim2008}. Curiously, no magnetic reflections were found at higher temperatures and up to 18.1\,K, where $\mu$SR sensed long-range magnetic order~\cite{Zheng2005ClinoPRL}. From the flat band of excitations observed for deuterated clinoatacamite powder by means of inelastic neutron scattering at 1.5\,K, in Ref.~\cite{Lee2007} it was concluded on a valence-bond solid ground state.
\bibitem{Wills2004} A. S. Wills, S. Raymond, and J.-Y. Henry, Magnetic ordering in a distorted $S=\frac{1}{2}$ pyrochlore antiferromagnet, \href{https://doi.org/10.1016/j.jmmm.2003.12.363}{J. Magn. Magn. Mater. \textbf{272--276}, 850 (2004)}.
\bibitem{Zheng2005Clino} X. G. Zheng, T. Kawae, Y. Kashitani, C. S. Li, N. Tateiwa, K. Takeda, H. Yamada, C. N. Xu, and Y. Ren, Unconventional magnetic transitions in the mineral clinoatacamite Cu$_{2}$Cl(OH)$_{3}$, \href{https://doi.org/10.1103/PhysRevB.71.052409}{Phys. Rev. B \textbf{71}, 052409 (2005)}.
\bibitem{Sokolov2019} D. A. Sokolov, N. Kikugawa, T. Helm, H. Borrmann, U. Burkhardt, R. Cubitt, J. S. White, E. Ressouche, M. Bleuel, K. Kummer, A. P. Mackenzie, and U. K. R\"o{\ss}ler, Metamagnetic texture in a polar antiferromagnet, \href{https://doi.org/10.1038/s41567-019-0501-0}{Nat. Phys. \textbf{15}, 671 (2019)}. 

\bibitem{Grice1996} J. D. Grice, J. T. Szyma\'nski, and J. L. Jambor, The crystal structure of clinoatacamite, a new polymorph of Cu$_{2}$Cl(OH)$_{3}$, Can. Mineral. \textbf{34}, 73 (1996).
\bibitem{Malcherek2009} T. Malcherek and J. Schl\"uter, Structures of the pseudo-trigonal polymorphs of Cu$_2$(OH)$_3$Cl, \href{https://dx.doi.org/10.1107/S0108768109013901}{Acta Cryst. B \textbf{65}, 334 (2009)}.
\bibitem{SI} See Supplemental Material at (URL will be inserted by publisher) for additional information on the structural characterization of clinoatacamite, the derivation of the motif of exchange couplings, the thermodynamic measurements, the $\mu$SR data analysis, the neutron diffraction experiments and the magnetic structure determination as well as the discussion of the phenomenological magnetic free energy. The Supplemental Material contains Refs.~\cite{Heinze2018,Malcherek2017,Guterding2016,Iqbal2015,Metropolis1953,Hastings1970,AlzateCardona2019,Kadena1994,FALCONpaper}.
\bibitem{Heinze2018} L. Heinze, R. Beltran-Rodriguez, G. Bastien, A. U. B. Wolter, M. Reehuis, J.-U. Hoffmann, K. C. Rule, and S. S\"ullow, The magnetic properties of single-crystalline atacamite, Cu$_2$Cl(OH)$_3$, \href{https://doi.org/10.1016/j.physb.2017.09.073}{Physica B \textbf{536}, 377 (2018)}.
\bibitem{Malcherek2017} T. Malcherek, B. Mihailova, and M. D. Welch, Structural phase transitions of clinoatacamite and the dynamic Jahn-Teller effect, \href{https://doi.org/10.1007/s00269-016-0859-9}{Phys. Chem. Minerals \textbf{44}, 307 (2017)}.
\bibitem{Guterding2016} D. Guterding, R. Valentí, and H. O. Jeschke, Reduction of magnetic interlayer coupling in barlowite through isoelectronic substitution, \href{https://doi.org/10.1103/PhysRevB.94.125136}{Phys. Rev. B \textbf{94}, 125136 (2016)}.
\bibitem{Iqbal2015} Y. Iqbal, H. O. Jeschke, J. Reuther, R. Valentí, I. I. Mazin, M. Greiter, and R. Thomale, Paramagnetism in the kagome compounds (Zn,Mg,Cd)Cu$_3$(OH)$_6$Cl$_{12}$, \href{https://doi.org/10.1103/PhysRevB.92.220404}{Phys. Rev. B \textbf{92}, 220404(R) (2015)}.
\bibitem{Metropolis1953} N. Metropolis, A. W. Rosenbluth, M. N. Rosenbluth, A. H. Teller, and E. Teller, Equation of State Calculations by Fast Computing Machines, \href{https://doi.org/10.1063/1.1699114}{J. Chem. Phys. \textbf{21}, 1087 (1953)}.
\bibitem{Hastings1970} W. K. Hastings, Monte Carlo sampling methods using Markov chains and their applications, \href{https://doi.org/10.1093/biomet/57.1.97}{Biometrika \textbf{57}, 97 (1970)}.
\bibitem{AlzateCardona2019} J. D. Alzate-Cardona, D. Sabogal-Suárez, R. F. L. Evans, and E. Restrepo-Parra, Optimal phase space sampling for Monte Carlo simulations of Heisenberg spin systems, \href{https://doi.org/10.1088/1361-648X/aaf852}{J. Phys.: Condens. Matter \textbf{31} 095802 (2019)}.
\bibitem{Kadena1994} Y. Kadena, and J. Mori, An effective Monte Carlo simulation on classical spin systems, \href{https://doi.org/10.1016/0375-9601(94)90762-5}{Phys. Lett. A \textbf{190}, 323 (1994).}
\bibitem{FALCONpaper} G. N. Iles, and S. Schorr, The HZB neutron Laue diffractometer: From E11 to FALCON, \href{https://doi.org/10.1080/10448632.2014.902701}{Neutron News \textbf{25}, 27 (2014)}.

\bibitem{Koepernik1999} K. Koepernik and H. Eschrig, Full-potential nonorthogonal local-orbital minimum-basis band-structure scheme, \href{https://doi.org/10.1103/PhysRevB.59.1743}{Phys. Rev. B \textbf{59}, 1743 (1999)}.
\bibitem{Perdew1996} J. P. Perdew, K. Burke, and M. Ernzerhof, Generalized Gradient Approximation Made Simple, \href{https://doi.org/10.1103/PhysRevLett.77.3865}{Phys. Rev. Lett. \textbf{77}, 3865 (1996)}.
\bibitem{Liechtenstein1995} A. I. Liechtenstein, V. I. Anisimov, and J. Zaanen, Density-functional theory and strong interactions: Orbital ordering in Mott-Hubbard insulators, \href{https://doi.org/10.1103/PhysRevB.52.R5467}{Phys. Rev. B \textbf{52}, R5467(R) (1995)}.
\bibitem{Xtal3.4} S. R. Hall, G. S. D. King, J. M. Stewart, Eds., \textit{Xtal} 3.4 User's Manual. University of Australia: Lamb, Perth (1995).
\bibitem{Sears1995} V. F. Sears, in \textit{International Tables for Crystallography}, edited by A. J. C. Wilson (Kluwer Academic Publishers, Dordrecht/Boston/London, 1995), Vol. C, p. 383.
\bibitem{Sears1992_absorption} V. F. Sears, Neutron scattering lengths and cross sections, \href{https://doi.org/10.1080/10448639208218770}{Neutron News \textbf{3}, 26 (1992)}.
\bibitem{Zachariasen1967} W. H. Zachariasen, A general theory of X-ray diffraction in crystals, \href{https://doi.org/10.1107/S0365110X67003202}{Acta Cryst. \textbf{23}, 558 (1967)}.
\bibitem{comm:trimermodel} Note that this motif of exchange couplings in the kagome plane was formerly studied in Ref.~\cite{Harris2013}, a study motivated by the coupling motif of clinoatacamite, but where the case $J_6 \gg J_4, J_5$ (our notation) was treated.
\bibitem{Harris2013} A. B. Harris and T. Yildirim, Spin dynamics of trimers on a distorted kagome lattice, \href{https://doi.org/10.1103/PhysRevB.88.014411}{Phys. Rev. B 88, 014411 (2013)}.
\bibitem{Khatami2012} E. Khatami, J. S. Helton, and M. Rigol, Numerical study of the thermodynamics of clinoatacamite, \href{https://doi.org/10.1103/PhysRevB.85.064401}{Phys. Rev. B \textbf{85}, 064401 (2012)}.
\bibitem{Jeschke2013} H. O. Jeschke, F. Salvat-Pujol, and R. Valentí, First-principles determination of Heisenberg Hamiltonian parameters for the spin-$1/2$ kagome antiferromagnet ZnCu$_3$(OH)$_6$Cl$_2$, \href{https://doi.org/10.1103/PhysRevB.88.075106}{Phys. Rev. B \textbf{88}, 075106 (2013)}.
\bibitem{Jeschke2015} H. O. Jeschke, F. Salvat-Pujol, E. Gati, N. H. Hoang, B. Wolf, M. Lang, J. A. Schlueter, and R. Valentí, Barlowite as a canted antiferromagnet: Theory and experiment, \href{https://doi.org/10.1103/PhysRevB.92.094417}{Phys. Rev. B \textbf{92}, 094417 (2015)}.

\bibitem{GPSpaper} A. Amato, H. Luetkens, K. Sedlak, A. Stoykov, R. Scheuermann, M. Elender, A. Raselli, D. Graf, The new versatile general purpose surface-muon instrument (GPS) based on silicon photomultipliers for \textmu{SR} measurements on a continuous-wave beam, \href{https://doi.org/10.1063/1.4986045}{Rev. Sci. Instrum. \textbf{88}, 093301 (2017)}.
\bibitem{Musrfit} A. Suter, and B. M. Wojek, Musrfit: A Free Platform-Independent Framework for $\mu$SR Data Analysis, \href{https://doi.org/10.1016/j.phpro.2012.04.042}{Phys. Proc. \textbf{30}, 69 (2012)}.

\bibitem{ILL_5-41-1083} L. St\"odter, K. Beauvois, B. Ouladdiaf, M. Reehuis, T. Reimann, K. C. Rule, S. S\"ullow, and A. U. B. Wolter, Institut Laue-Langevin (ILL), \url{https://doi.org/10.5291/ILL-DATA.5-41-1083} (2021).
\bibitem{ILL_EASY-1612} E. Chan, Institut Laue-Langevin (ILL), \url{https://doi.org/10.5291/ILL-DATA.EASY-1612} (2025).
\bibitem{ILL-5-41-1349} L. St\"odter, J. Avtarovski, E. Chan, C. Kastner, M. Reehuis, K. C. Rule, S. S\"ullow, and A. U. B. Wolter, Institut Laue-Langevin (ILL), \url{https://doi.ill.fr/10.5291/ILL-DATA.5-41-1349} (2026).
\bibitem{E2paper} Helmholtz-Zentrum Berlin f\"ur Materialien und Energie, E2: The Flat-Cone Diffractometer at BER II, \href{https://doi.org/10.17815/jlsrf-4-110}{J. Large-Scale Res. Facil. \textbf{4}, A129 (2018)}.
\bibitem{Zheng2024} X.-G. Zheng, M. Hagihala, I. Yamauchi, E. Nishibori, T. Honda, T. Yuasa, and C.-N. Xu, A novel Kagome uud-ddu spin order in Heisenberg spin-$1/2$ Clinoatacamite Cu$_4$(OH)$_6$Cl$_2$, the parent compound of Herbertsmithite, \href{https://doi.org/10.48550/arXiv.2411.01277}{arXiv:2411.01277} [cond-mat.str-el].

\bibitem{Bertaut1968} E. F. Bertaut, Representation analysis of magnetic structures, \href{https://doi.org/10.1107/S0567739468000306}{Acta Cryst. A \textbf{24}, 217 (1968)}.
\bibitem{Fullprof} J. Rodr\'iguez-Carvajal, Recent advances in magnetic structure determination by neutron powder diffraction, \href{https://doi.org/10.1016/0921-4526(93)90108-I}{Physica B \textbf{192}, 55 (1993)}.
\bibitem{Brown1995} P. J. Brown, in \textit{International Tables for Crystallography}, edited by A. J. C. Wilson (Kluwer Academic Publishers, Dordrecht/Boston/London, 1995), Vol. C, p. 391.
\bibitem{comm:Cu1Cu2} During the refinement we treated Cu(1)/Cu(2) as one magnetic sublattice as they have the same point symmetry $\overline{1}$.

\bibitem{ChaikinLubensky1995} P. M. Chaikin and T. C. Lubensky, Principles of Condensed Matter Physics, Cambridge UP, Cambridge UK (1995).
\bibitem{Dzyaloshinskii1964} I. E. Dzyaloshinskii, Sov.\ Phys.\ JETP {\textbf{19}}, 960 (1964).
\bibitem{Schweika2022} W. Schweika, M. Valldor, J. D. Reim, and U. K. R\"o{\ss}ler, Chiral spin liquid ground state in YBaCo$_3$FeO$_7$, \href{https://doi.org/10.1103/PhysRevX.12.021029}{Phys. Rev. X \textbf{12}, 021029 (2022)}.

\bibitem{VESTA} K. Momma and F. Izumi, VESTA 3 for three-dimensional visualization of crystal, volumetric and morphology data, \href{https://doi.org/10.1107/S0021889811038970}{J. Appl. Cryst. \textbf{44}, 1272 (2011)}.
                                                      
\end{thebibliography}
\end{document}